\documentclass[a4paper,reprint,nofootinbib]{revtex4-1}

\pdfoutput=1

\usepackage{amsmath,amssymb}
\usepackage[squaren]{SIunits}
\usepackage{graphicx}
\usepackage[style=american]{csquotes}

\newcommand{\reffig}[1]{FIG.~\ref{#1}}
\newcommand{\refeq}[1]{Eq.~(\ref{#1})}
\newcommand{\refsec}[1]{Section~\ref{#1}}

\newcommand{\tp}[1]{\ensuremath{^3P_{#1}}}

\newcommand{\nei}[1]{\ensuremath{\mathrm{^{#1}Ne}}}

\newcommand{\dd}[1]{\ensuremath{\mathrm{d}{#1}}}

\newcommand{\um}{\micro\meter}
\newcommand{\mm}{\milli\meter}
\newcommand{\cm}{\centi\meter}
\newcommand{\ms}{\milli\second}
\newcommand{\mus}{\micro\second}
\newcommand{\mps}{\meter\per\second}

\begin{document}

\title{Multistage Zeeman deceleration of metastable neon}
\author{Alex W. Wiederkehr}
\author{Michael Motsch}
\author{Stephen D. Hogan}
\author{Markus Andrist}
\author{Hansj\"urg Schmutz}
\author{Bruno Lambillotte}
\author{Josef A. Agner}
\author{Fr\'ed\'eric Merkt}
\affiliation{Laboratorium f\"ur Physikalische Chemie, ETH Z\"urich, CH-8093, Switzerland}

\begin{abstract}
A supersonic beam of metastable neon atoms has been decelerated by exploiting the interaction between the magnetic moment of the atoms and time-dependent inhomogeneous magnetic fields in a multistage Zeeman decelerator. Using 91 deceleration solenoids, the atoms were decelerated from an initial velocity of \unit{580}{\mps} to final velocities as low as \unit{105}{\mps}, corresponding to a removal of more than $95\,\%$ of their initial kinetic energy. The phase-space distribution of the cold, decelerated atoms was characterized by time-of-flight and imaging measurements, from which a temperature of \unit{10}{\milli\kelvin} was obtained in the moving frame of the decelerated sample. In combination with particle-trajectory simulations, these measurements allowed the phase-space acceptance of the decelerator to be quantified. The degree of isotope separation that can be achieved by multistage Zeeman deceleration was also studied by performing experiments with pulse sequences generated for \nei{20} and \nei{22}.
\end{abstract}

\maketitle

\section{Introduction}
\label{sec:intro}

The development of methods to produce atomic and molecular samples with precisely controlled translational and internal degrees of freedom has received considerable attention in recent years, with particular emphasis on the generation of cold ($\mathrm{T} = \unit{1}{\milli\kelvin} - \unit{1}{\kelvin}$) and ultracold ($\mathrm{T} < \unit{1}{\milli\kelvin}$) samples. This interest in the study of cold molecules is motivated by applications in a broad range of disciplines~\cite{bell09b, carr09a}, including the study of collision processes and chemical reactions in a temperature regime where quantum-mechanical phenomena may dominate the reaction dynamics, high-precision spectroscopy, quantum information processing and quantum simulations, and the study of exotic quantum phases arising from the long-range anisotropic dipole-dipole interaction between polar molecules.

Amongst the methods developed for the production of cold molecules (for an overview, see, for example, \cite{doyle04a, dulieu06a, carr09a, heiner06a, vandemeerakker08a, bell09b, schnell09a, hogan11a, smith08a, krems09a} and references therein), those based on the deceleration of supersonic molecular beams are particularly suited to collision experiments~\cite{gilijamse06a, sawyer08a, vandemeerakker09a, kirste10a, sawyer11a, scharfenberg10a, parazzoli11a}: The final velocity of the decelerated particles is tunable over a wide range with a narrow velocity spread~\cite{bethlem99a, scharfenberg10a}, the deceleration process is internal-state selective~\cite{vandemeerakker06a, hoekstra07b}, and the initial phase-space density is preserved by operating the decelerator in a phase-stable manner~\cite{veksler45a, mcmillan45a, humphries86a, bethlem00a, bethlem02a, wiederkehr10b}. Depending on the characteristics of the species of interest, multistage Stark deceleration~\cite{bethlem99a}, Rydberg-Stark deceleration~\cite{vliegen05a, hogan09a}, multistage Zeeman deceleration~\cite{vanhaecke07a, narevicius08a, lavertofir11b, trimeche11a, lavertofir11a}, or optical deceleration~\cite{fulton04a} may be best suited.

Multistage Zeeman deceleration relies on the interaction between paramagnetic species and time-dependent inhomogeneous magnetic fields, and is therefore particularly attractive for open-shell systems such as molecular radicals and metastable atoms and molecules. In this article, we present results on the deceleration of a supersonic beam of neon atoms in the metastable $(1s)^2 (2s)^2 (2p)^5 (3s)^1$ \tp{2} state using a 91-stage Zeeman decelerator. Because of their favorable ratio of magnetic moment to mass and the straightforward detection of beams of metastable atoms, light metastable rare-gas atoms are well suited to multistage Zeeman deceleration, and beams of such atoms have been used by several groups to characterize their Zeeman decelerators~\cite{narevicius08a, lavertofir11b, trimeche11a, lavertofir11a}. Moreover, cold decelerated beams of metastable rare-gas atoms are of interest in the study of Penning ionization processes~\cite{siska93a, feltgen99a} with a high energy resolution at low collision energies~\cite{toennies07a, chandler10a}. Metastable neon has been laser cooled and trapped~\cite{shimizu89a}. Multistage Zeeman deceleration also provides a means to prepare atomic or molecular samples in a selected magnetic sublevel, which may be used in collision experiments and complement existing techniques to prepare aligned and oriented samples (see~\cite{zare98a, bartlett08a} and references therein).

This article describes the extension of our modular high-repetition-rate Zeeman decelerator design, which has previously been employed for deceleration and trapping of hydrogen and deuterium atoms~\cite{hogan08c, hogan08d, wiederkehr10a, wiederkehr10b}, to the deceleration of heavier species, and also presents the characterization of the velocity distribution of the decelerated samples. Characteristic features of this decelerator, next to its modular nature, are its high repetition rate (\unit{10}{\hertz}) and the solenoid design which enables one to easily calculate magnetic field distributions. The latter advantage is of crucial importance for particle trajectory simulations and their use in the optimization of the deceleration process. An introduction into design considerations relevant to multistage Zeeman deceleration will be given first, followed by a detailed description of the individual deceleration stages, their assembly into modules, and the integration of the modules into the multistage decelerator. The electronics circuits developed and used to pulse high currents through the deceleration solenoids with short rise and fall times are also described.

\section{Experimental Setup}
\label{sec:setup}

\begin{figure*}
\includegraphics{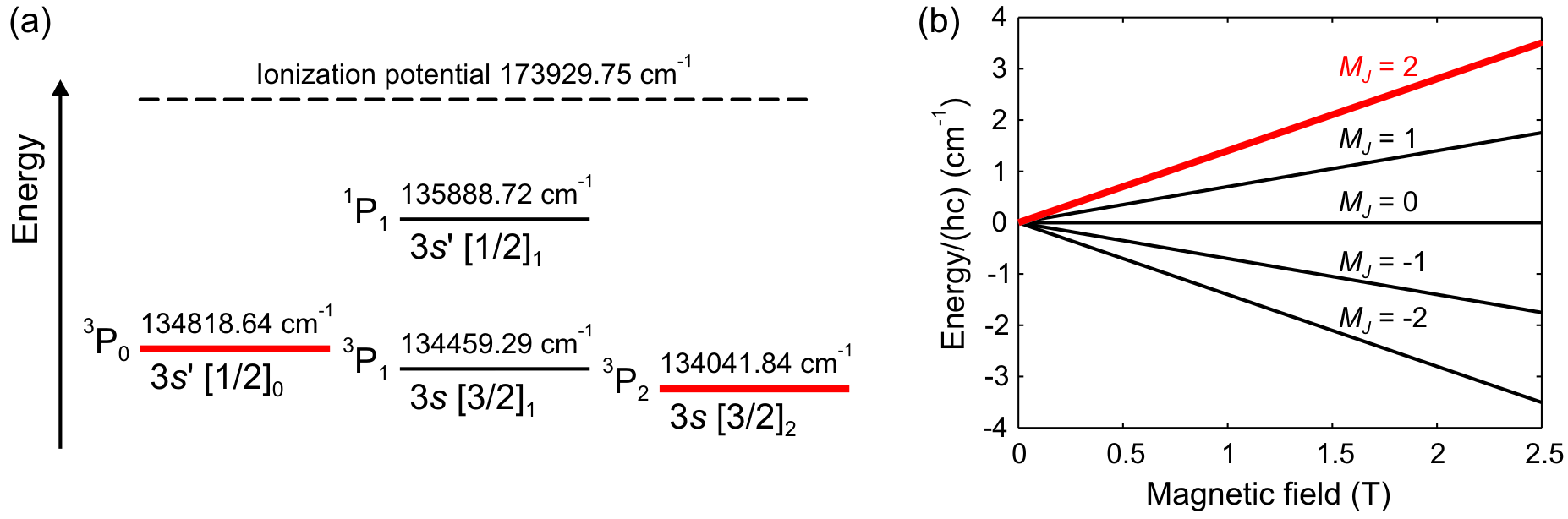}
\caption{\label{fig:energy_levels}
(a) Schematic energy level diagram of the states of Ne relevant to the experiment, with wave numbers given relative to the $(2p)^6$ $^1S_0$ ground  state~\cite{nistatomicdatabase}. The metastable states \tp{2}\ and \tp{0}\ are indicated by thick red lines. (b) Zeeman effect in the \tp{2} state. The low-field-seeking state $M_J=2$ with the largest magnetic moment is depicted in red.
}
\end{figure*}

In a multistage Zeeman decelerator, the longitudinal velocity of low-field-seeking particles in a supersonic beam is reduced by converting kinetic energy into potential energy through repeated application of pulsed inhomogeneous magnetic fields~\cite{hogan11a}. Typical laboratory magnetic fields, which can be reliably produced and rapidly switched with conventional solenoids, are on the order of a few Tesla. These give rise to Zeeman shifts $\Delta W_\mathrm{Z}/(h c)$ \unit{\approx 1}{\reciprocal\cm} for a magnetic moment of one Bohr magneton ($\mu_\mathrm{B}$). To give an example, the kinetic energy of a particle of mass number $N_\mathrm{M}$ propagating at a velocity of \unit{400}{m/s} amounts to $E_\mathrm{kin}/(hc)\approx N_\mathrm{M} \times\unit{6.7}{\reciprocal\cm}$. Many successive deceleration stages are therefore needed to bring even small atomic or molecular systems to a standstill. Moreover, a phase-stable operation of the decelerator is essential to maintain the initial phase-space density while decelerating parts of the supersonic beam to the desired final velocity~\cite{bethlem02a, wiederkehr10b, hogan11a}.

A simplified energy level diagram of the metastable states of neon relevant to the experiment is shown in \reffig{fig:energy_levels}. The Zeeman shift of the metastable \tp{2} state is $\Delta W_\mathrm{Z} = g_J \mu_\mathrm{B} M_J B$, with $B$ the magnetic field and $g_J=1.50$~\cite{lurio60a}. At a magnetic field of \unit{1.7}{\tesla}, reached in our decelerator at a phase angle of \unit{45}{\degree}, the low-field-seeking state $M_J=2$ used in the deceleration experiments is subject to a Zeeman shift of about \unit{2.4}{\reciprocal\cm}.

\subsection{Overview of the experimental setup}
\label{sec:setup:overview}

\begin{figure*}
\includegraphics{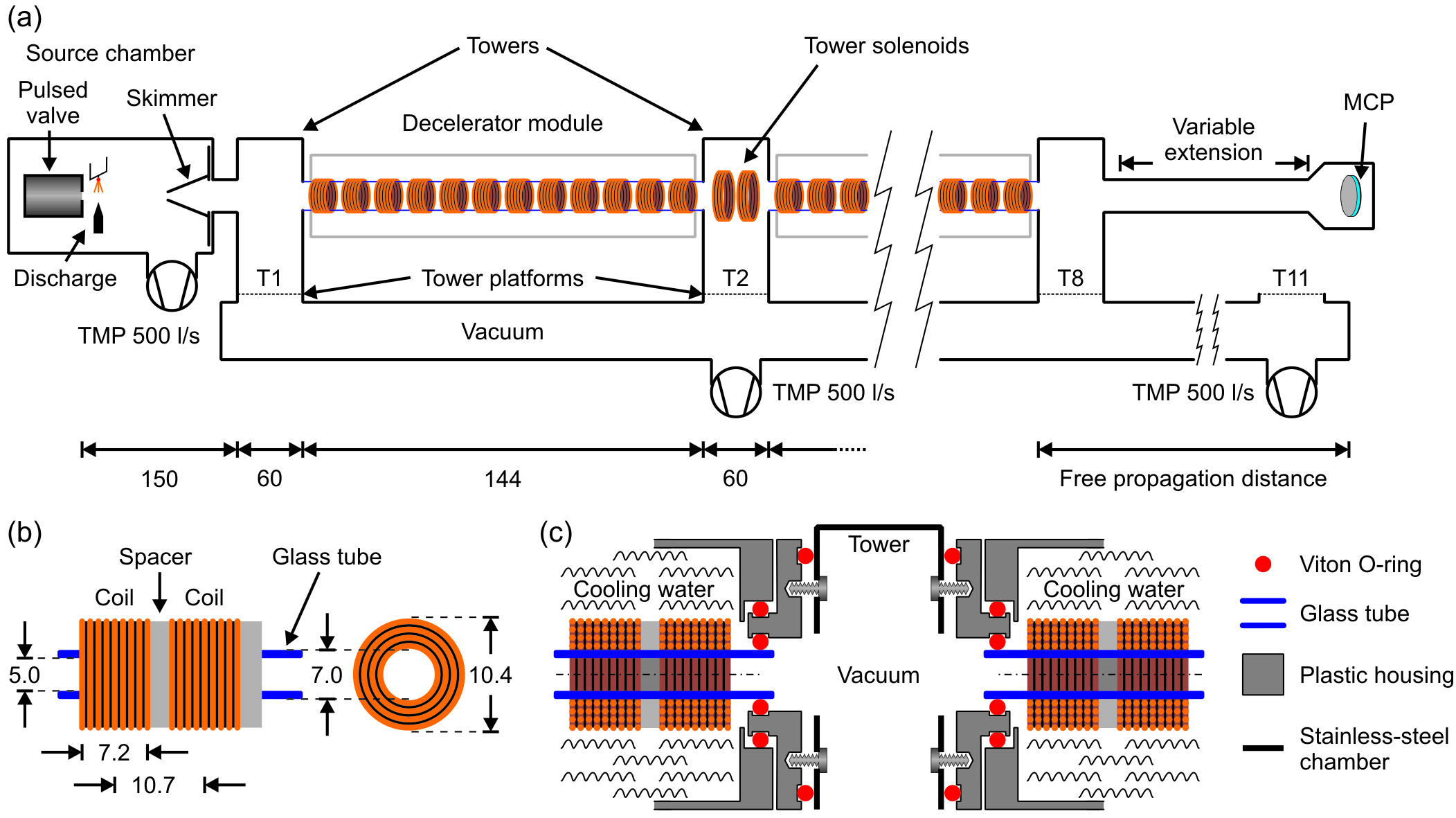}
\caption{\label{fig:exp_setup}
Schematic representation (not to scale) of the experimental setup. All distances are given in mm.
(a)~The modular decelerator with its main components: the metastable atom source, the evacuated volume with the tower platforms T1--T11, the towers, and the deceleration modules. The metastable atoms are detected after an adjustable free-propagation distance to the micro-channel-plate (MCP).
(b)~Dimensions of the deceleration solenoids and their assembly in a deceleration module.
(c)~Interface between the water-cooled deceleration modules and the pumping towers. The solenoids in the tower are not included for clarity.
}
\end{figure*}

A schematic representation of the experimental setup is shown in \reffig{fig:exp_setup}~(a). The main components are a pulsed source producing a supersonic beam of metastable neon atoms, the modular Zeeman decelerator, and the electronics driving the current through the solenoids to generate strong magnetic fields for deceleration. In the design of the decelerator and its electronics, the following aspects have been considered to make the experiment flexible and versatile:

\begin{enumerate}
\item
The decelerator was set up in a modular way to enable the deceleration of different atomic or molecular systems. The number of deceleration stages required to fully decelerate a given species depends on its magnetic-moment-to-mass ratio. For example, 12 deceleration stages are needed for H~\cite{hogan08c}, 24 deceleration stages for D~\cite{wiederkehr10a}, and around 90 deceleration stages for metastable Ne.
\item
Collision and high-resolution spectroscopic experiments with pulsed supersonic beams are typically performed at repetition rates between \unit{10-30}{\hertz}. Ideally, the repetition rate of the decelerator should lie in this range, which places strong constraints on the cooling of the solenoids.
\item
To keep the number of deceleration stages low, the kinetic energy removed at each deceleration stage should be as large as possible.
\item
To avoid particle loss by nonadiabatic transitions at low field, a quantization field must be maintained along the entire trajectories of the decelerated particles~\cite{hogan07a}.
\item
The deceleration efficiency benefits from precise knowledge of the spatial and temporal shapes of the magnetic field pulses. In addition, \enquote{clean} magnetic-field pulses simplify the calculation of deceleration pulse sequences.
Metallic materials close to the solenoids should therefore be avoided to minimize the influence of eddy currents. Moreover, high-permeability materials, which can be used for field-enhancement purposes~\cite{narevicius08a, narevicius08b}, result in a rest magnetization that is difficult to quantify.
\item
Whereas electronic devices for the fast switching of high voltages are commercially available, dedicated fast switching electronics for pulsing high currents through inductive loads are required.
\item
A high vacuum should be maintained in the deceleration beam line to avoid losses by collisions with the background gas and with particles not accepted by the decelerator~\cite{hogan11a}.
\end{enumerate}

The detection of metastable atoms or molecules in a beam experiment is straightforward: If the internal energy of the metastable state exceeds the work function of the micro-channel-plate (MCP) detector, an electron current is generated upon impact on the MCP.  With an internal energy of \unit{16.62}{\electronvolt}, the \tp{2}\ state of Ne clearly fulfills this requirement. Depending on the specific measurement, the MCPs used in the experiments described here were assembled in a standard chevron configuration for the measurement of time-of-flight (TOF) profiles, or additionally equipped with a phosphor screen for imaging the radial distribution of the decelerated atoms with a CCD camera. Compared to detection by photoionization with a pulsed laser, which was used in the Zeeman deceleration of H and D \cite{hogan07a}, this detection method has the advantage of giving a complete time-of-flight trace without the need of scanning the delay time of the laser with respect to the nozzle opening time. Hence, data acquisition is fast, and a wide parameter range can be sampled in a short time. This advantage is particularly important during the initial optimization of the decelerator and the deceleration pulse sequences.

\subsection{Metastable-neon source}
\label{sec:setup:source}

Neon is expanded from a reservoir at a stagnation pressure of \unit{5}{\bbar} into vacuum through the \unit{250}{\um} diameter orifice of a pulsed Even-Lavie valve~\cite{even00a}. To increase the density in the beam and reduce the initial speed of the atoms, the body of the valve is thermally connected to a liquid-nitrogen reservoir by copper braids and held at a temperature of about \unit{125}{\kelvin}. To populate the metastable \tp{2}\ state, a discharge is operated in the expansion region~\cite{halfmann00a}. The distance between the nozzle orifice and the discharge is kept as short as possible to produce the metastable atoms in a high-density region, which was found to be efficient and also cools the beam in the expansion. For this purpose, a thin metal tip, to which a DC voltage of \unit{+180}{\volt} is applied, is placed \unit{1}{\mm} away from the ceramic front plate of the valve body, with the grounded metal poppet of the valve acting as the second discharge electrode. The tip is radially displaced from the symmetry axis of the nozzle to avoid obstruction of the supersonic beam. To efficiently operate the discharge at this low voltage and minimize heating of the beam, the discharge is seeded with thermal electrons continuously emitted from a filament (Agar Scientific A054C, \unit{0.2}{\mm} tungsten wire, AC current up to \unit{6}{\ampere} rms) placed downstream from the nozzle.

To characterize the supersonic beam, we have recorded the time-of-flight distribution of the metastable neon atoms impinging on the MCP at the exit of the decelerator, about \unit{1.8}{\meter} from the nozzle. From such time-of-flight measurements, the mean velocity of the metastable neon atoms was determined to be \unit{580-590}{\mps}. Further properties of the discharge source can be determined by threshold photoionization spectroscopy using a pulsed dye laser. After propagation through the decelerator, only the metastable \tp{2}\ and \tp{0}\ states are expected to be populated in the beam. Scanning the laser over their ionization thresholds at \unit{39111.11}{\reciprocal{\cm}} (\tp{0}) and \unit{39887.91}{\reciprocal{\cm}} (\tp{2})~\cite{nistatomicdatabase}, and recording the ion yield, we found a ratio of population in the \tp{2} state to that in the \tp{0} state of 5:1, which corresponds to the degeneracy factors of these energy levels. Because the state \tp{2}, $M_J=2$ has the largest magnetic moment it is the most convenient state for deceleration, and the deceleration pulse sequences were optimized for its properties. Assumining an equal initial population of all magnetic sublevels, the atoms in the $M_J=2$ state represent only one sixth of the population of metastable atoms. In the measurements, neon gas with natural isotopic composition was used, the most abundant isotopes being \nei{20} (90.5\,\%) and \nei{22} (9.3\,\%). In most experiments, the deceleration pulse sequences were generated for \nei{20}. The occurrence of two different isotopes in the beam, however, offers the possibility of investigating the selectivity of multistage Zeeman deceleration on the magnetic-moment-to-mass ratio. This was done by applying deceleration pulse sequences generated for \nei{20} and \nei{22}, and detecting the cold decelerated atoms in a mass-selective manner.

\subsection{Multistage Zeeman decelerator}
\label{sec:setup:decel}

The deceleration solenoids are at the heart of the multistage Zeeman decelerator. In their design and optimization, the following factors have been considered.
\begin{enumerate}
\item
The open area of the decelerator, determined by the free inner diameter of the solenoids, should be matched to the  profile of the supersonic beam. The dimensions are therefore similar to those of the electrodes in a multistage Stark decelerator.
\item
The solenoids should produce a large decelerating field gradient on axis and provide transverse focusing of the beam.
\item
The length of the solenoids should be kept as short as possible to maximise the energy removed per unit length.
\item
The length of the solenoids should be matched to the initial speed of the supersonic beam and to the achievable switching times of the magnetic deceleration fields. For efficient phase-stable deceleration, the beam should not penetrate too far into the solenoids during the switch-off time of the field.
\end{enumerate}

The individual deceleration solenoids were made of enameled wire having a copper-core diameter of \unit{400}{\um}  (Elektrisola XHTW, ABP15, grade 1B). The wire was wound in four layers with 62 windings in total, with a free inner solenoid diameter of \unit{7}{\mm}, an outer diameter of \unit{10.4}{\mm}, and a length of \unit{7.2}{\mm}. The length of these solenoids was chosen as a compromise between achieving the maximum on-axis field for a given number of windings per unit length at this inner diameter while at the same time maximizing the deceleration that can be achieved per unit length. Electrically, the solenoids can be characterized by an inductance of \unit{21}{\micro\henry} and an ohmic resistance of about \unit{0.24}{\ohm} at room temperature. To improve the mechanical stability of the solenoids, a supporting structure is produced by curing the enamel coating.

\begin{figure}
\includegraphics{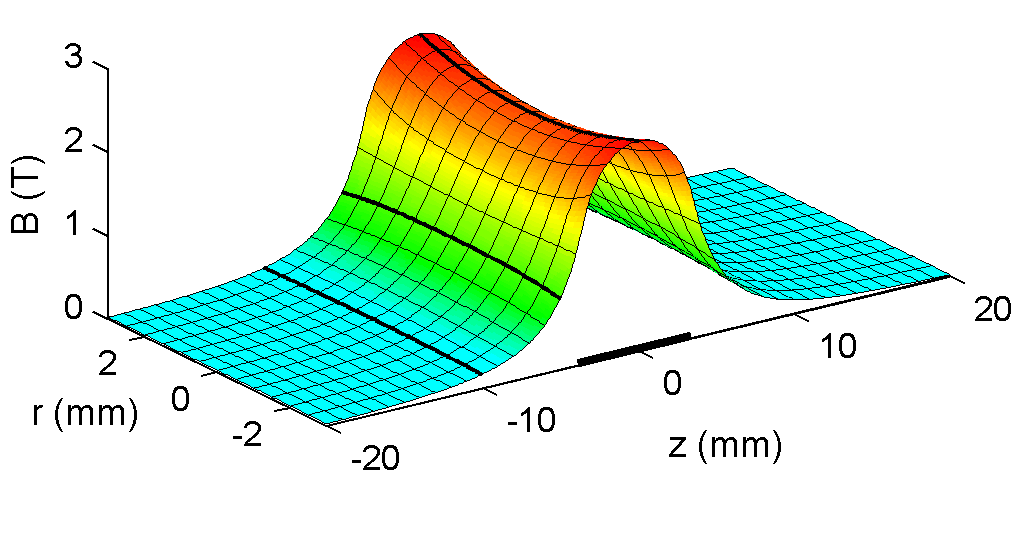}
\caption{\label{fig:field}
The axially symmetric magnetic field produced by the \unit{7.2}{\mm}-long deceleration solenoid, centered at z=0 and operated at a current of \unit{300}{\ampere}. On the symmetry axis of the solenoid, the barrier has a height of \unit{2.2}{\tesla}. The thick curves on the surface at \unit{z=0}{\mm}, \unit{z=-5}{\mm}, and \unit{z=-10}{\mm} highlight the concave shape of the field near the center of the solenoid, which gives rise to radial focusing forces, and the convex shape at larger distances from the center, where forces become radially defocusing. The longitudinal extension of the solenoid is indicated by the thick bar on the z-axis.
}
\end{figure}

The magnetic field distribution of a solenoid calculated for a typical current of \unit{300}{\ampere}, the maximum pulsed current rating of the IGBT (insulated-gate bipolar transistor) switches used, is depicted in \reffig{fig:field}. The solenoids create a magnetic field barrier of \unit{2.2}{\tesla} along the supersonic beam propagation axis, which is used for deceleration. Near the center of the solenoid, the field is concave, which gives rise to radial focusing forces. At larger distances from the center, the field is convex, and forces in the radial direction are defocusing. This alternation plays an important role in the transverse dynamics in the multistage Zeeman decelerator as discussed by Wiederkehr et al.~\cite{wiederkehr10b}.

When a current of \unit{300}{\ampere} is pulsed through a solenoid, an electrical power of \unit{20}{\kilo\watt} is dissipated. At a repetition rate of the decelerator of \unit{10}{\hertz} and a typical pulse duration of \unit{50}{\mus}, a heat load of \unit{10}{\watt} per solenoid must be removed by the cooling system. To efficiently cool the solenoids and simplify the construction, 12 individual solenoids are assembled into one module as shown in \reffig{fig:exp_setup}. Within such a module, the supersonic beam propagates inside a quartz glass tube (inner diameter \unit{5}{\mm}, wall thickness \unit{1}{\mm}) representing the evacuated beamline. The solenoids, separated by ceramic (BNP2) spacers at a center-to-center distance of \unit{10.7}{\mm}, are glued on the outside of this glass tube with a heat-conducting epoxy (Aremco-Bond 860). The epoxy prevents mechanical distortions when high currents are pulsed and protects the outermost layer of the solenoid from abrasion by the circulating cooling water. The glass tube assembled with the solenoids is then inserted into a plastic casing which also houses electrical feedthroughs to supply the current to the solenoids. This design allows the immersion of the solenoids in circulating cooling water to efficiently remove the heat, such that the decelerator can be operated at a high repetition rate of \unit{10}{\hertz}.

Before the supersonic beam enters the decelerator, it passes a \unit{2}{\mm} diameter skimmer placed \unit{10}{\cm} downstream from the nozzle. The nozzle is positioned in the source vacuum chamber with good optical access, which is necessary when the species to be decelerated are produced by either photoexcitation or photolysis from a precursor molecule. The size of the skimmer orifice is chosen so as to match the transverse profile of the supersonic beam to the radial acceptance of the decelerator. To reduce the gas load from the source in the decelerator and avoid losses by collisions with the background gas~\cite{hogan11a}, the beam first propagates through a tower extending above an evacuated volume (see \reffig{fig:exp_setup}). The large vacuum chamber, covering the full length of the decelerator, is evacuated by two \unit{500}{\litre\per\second} turbo molecular pumps to a pressure in the low \unit{10^{-7}}{\milli\bbar} range.

For the deceleration of Ne (\tp{2}), five modules of \unit{144}{\mm} length, each equipped with 12 solenoids, and one additional \unit{348}{\mm}-long module with 31 deceleration solenoids~\footnote{When the \unit{348}{\mm}-long module is used, one tower is skipped, and the length of the module is given by twice the tower spacing plus the tower length.} were used. As shown in \reffig{fig:exp_setup}, the modules are attached with viton O-ring seals to the towers. Therefore, modules can be added or removed easily to adapt the length of the decelerator to different species. The 12-solenoid modules are similar to the ones used for the deceleration and trapping of H and D atoms, and were designed so as to permit deceleration of H atoms to a standstill~\cite{hogan08c, hogan08d, wiederkehr10a, wiederkehr10b}.

Interrupting the sequence of deceleration stages by the \unit{60}{\mm}-long propagation distances through the towers is necessary to maintain a high vacuum in the decelerator, but it introduces additional technical difficulties. In the case of neon, only atoms in the $\tp{2}, M_J=2$ state are efficiently decelerated. To avoid nonadiabatic transitions of the low-field-seeking states near zero field, which would cause particle loss~\cite{hogan07a, hogan11a}, it is important to maintain a quantization field along the complete trajectory through the decelerator. Within the modules, this is achieved by switching on the field of the next solenoid just before the field of the active solenoid is switched off. In each tower, two larger solenoids (free inner solenoid radius \unit{6}{\mm}, six layers with 11 windings each, \unit{1}{\mm} diameter enameled copper tubing with a wall thickness of \unit{0.2}{\mm}, center-to-center spacing of the solenoids \unit{28}{\mm}), depicted in \reffig{fig:exp_setup}, define the quantization axis~\cite{wiederkehr10a}. The adjustment of the currents of these solenoids can also be used to manipulate the transverse motion of the atoms.

\subsection{Electronics driving the deceleration solenoids}
\label{sec:setup:electronics}

\begin{figure}
\includegraphics{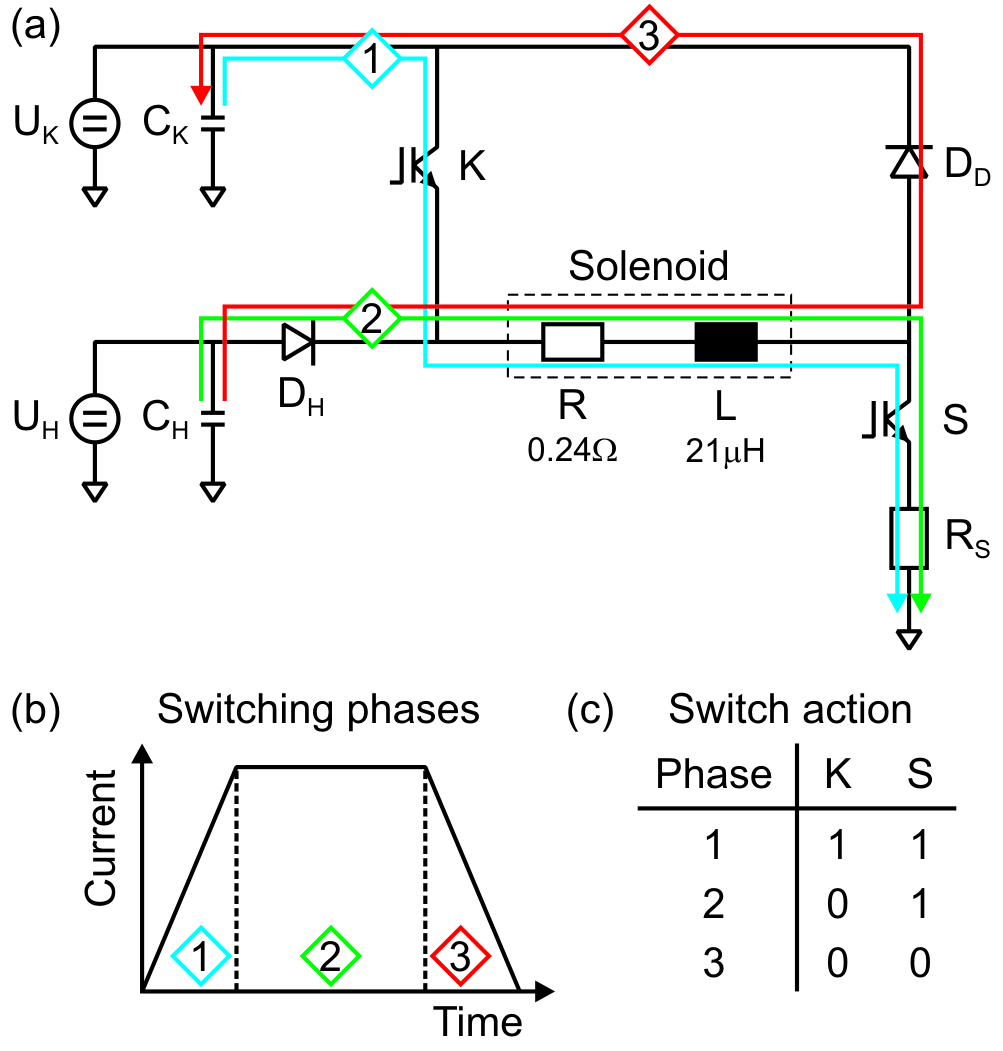}
\caption{\label{fig:electronics_scheme}
(a) Operation principle of the electronic circuit used to generate pulsed magnetic fields with short rise and fall times. The deceleration solenoid is represented by its equivalent resistance and inductance inside the dashed frame.
$\mathrm{U_K}$: kick voltage supply (\unit{800}{V});
$\mathrm{C_K}$: storage capacitor (\unit{50}{\micro\farad});
$\mathrm{U_H}$: hold voltage supply (\unit{100-120}{V});
$\mathrm{C_H}$: storage capacitor (\unit{26.4}{\milli\farad});
$\mathrm{D_H}$, $\mathrm{D_D}$: hold diode and decay diode;
S, K: IGBT (insulated-gate bipolar transistor) switches;
$\mathrm{R_S}$: shunt resistor (\unit{10}{\milli\ohm}).
The current flow during the switch-on (1), hold (2), and switch-off (3) phases is indicated by the blue, green, and red arrows next to the wires.
The nominal values of $\mathrm{C_K}$ and $\mathrm{C_H}$ are chosen to supply the current pulses to six deceleration solenoids (see \reffig{fig:electronics_module}).
(b) Schematic temporal profile of the current through the solenoid.
(c) Logical states of the IGBT switches during the three switching phases (0: open, no current flow; 1: closed).
}
\end{figure}

\begin{figure}
\includegraphics{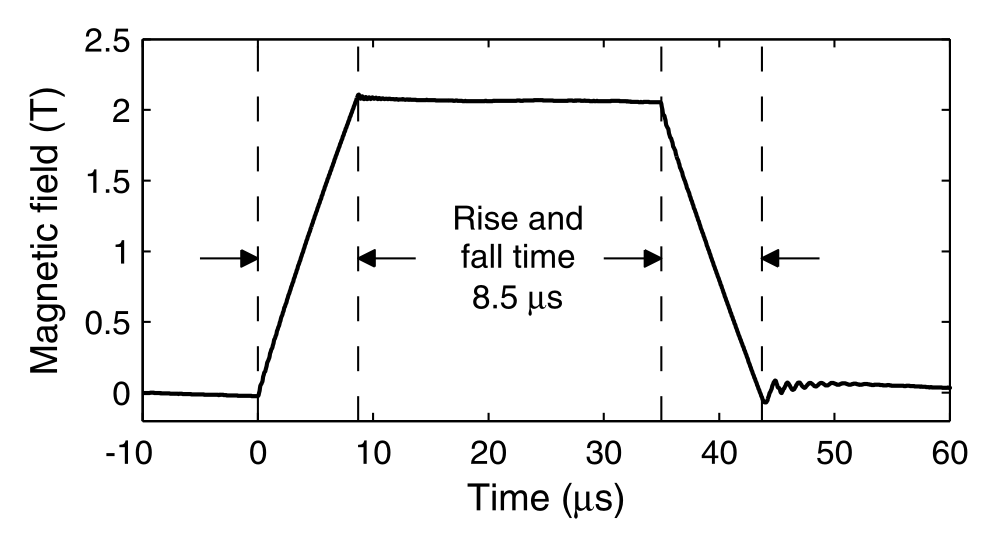}
\caption{\label{fig:fieldmeas}
Magnetic field at the center of a deceleration solenoid. The field was obtained by integrating the voltage induced across a pickup coil (\unit{1.4}{\mm} free inner diameter, 15 windings of a \unit{0.1}{\mm}-diameter enameled copper wire, L=\unit{250}{\nano\henry}) placed in the bore of the solenoid.
}
\end{figure}

\begin{figure*}
\includegraphics{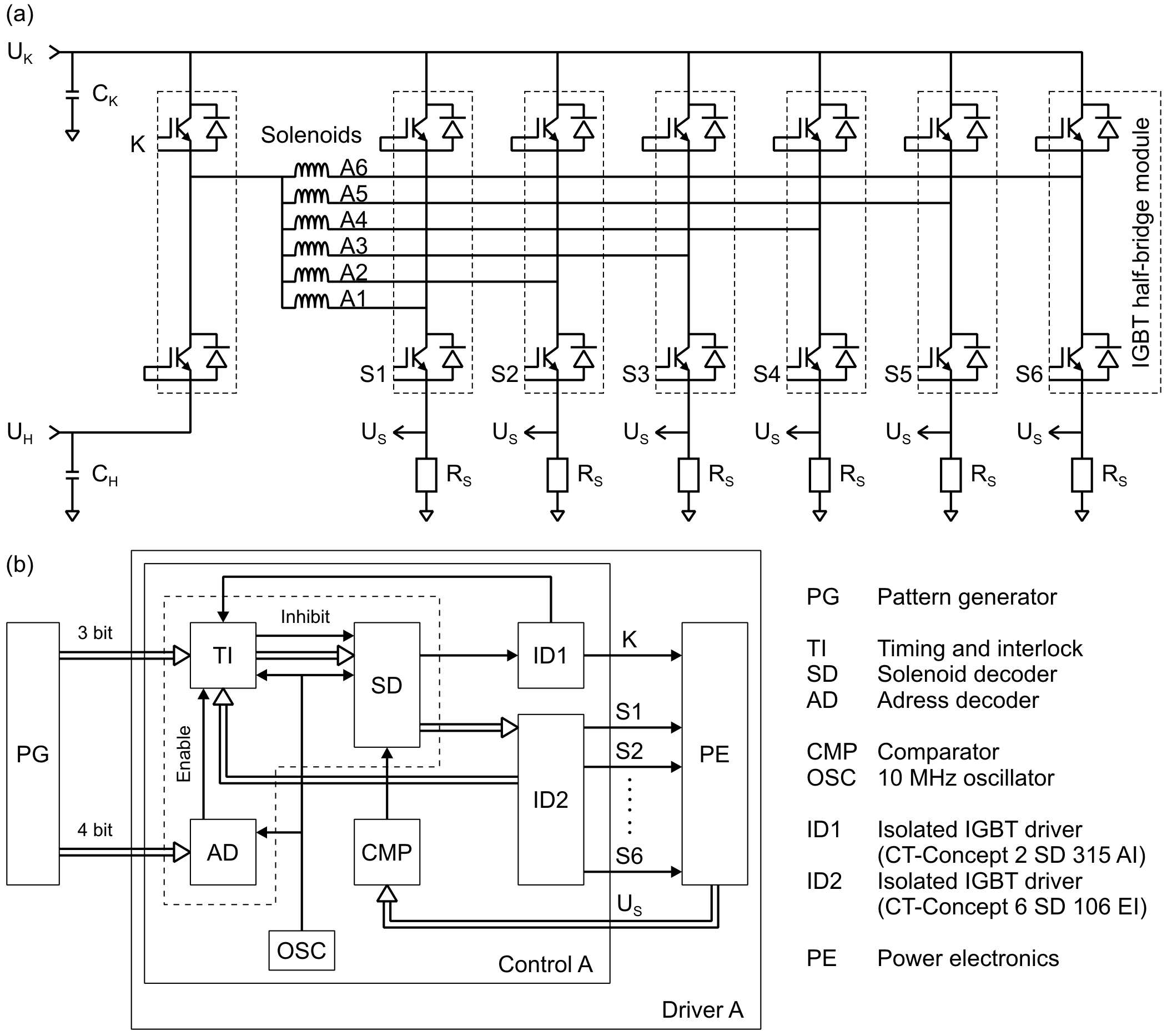}
\caption{\label{fig:electronics_module}
(a)~Schematic diagram of the power electronics driver A used for pulsing the current through deceleration solenoids A1--A6 in a deceleration module consisting of 12 solenoids.
$\mathrm{U_{K(H)}}$, $\mathrm{C_{K(H)}}$: kick (hold) voltage supply and storage capacitor as defined in \reffig{fig:electronics_scheme};
K: control input for kick voltage switch;
S1--S6: control input for the switches used to energize one of the deceleration solenoids A1--A6 at a time. The switch action corresponds to that of S in \reffig{fig:electronics_scheme}.
(b)~Schematic diagram of the interface and control circuitry for driver A. The circuitry bounded by the dashed line is realized as a synchronous design with CMOS programmable logic devices (CPLD, Lattice M4A5), resulting in a time resolution of \unit{100}{\nano\second}.
}
\end{figure*}

\begin{figure}
\includegraphics{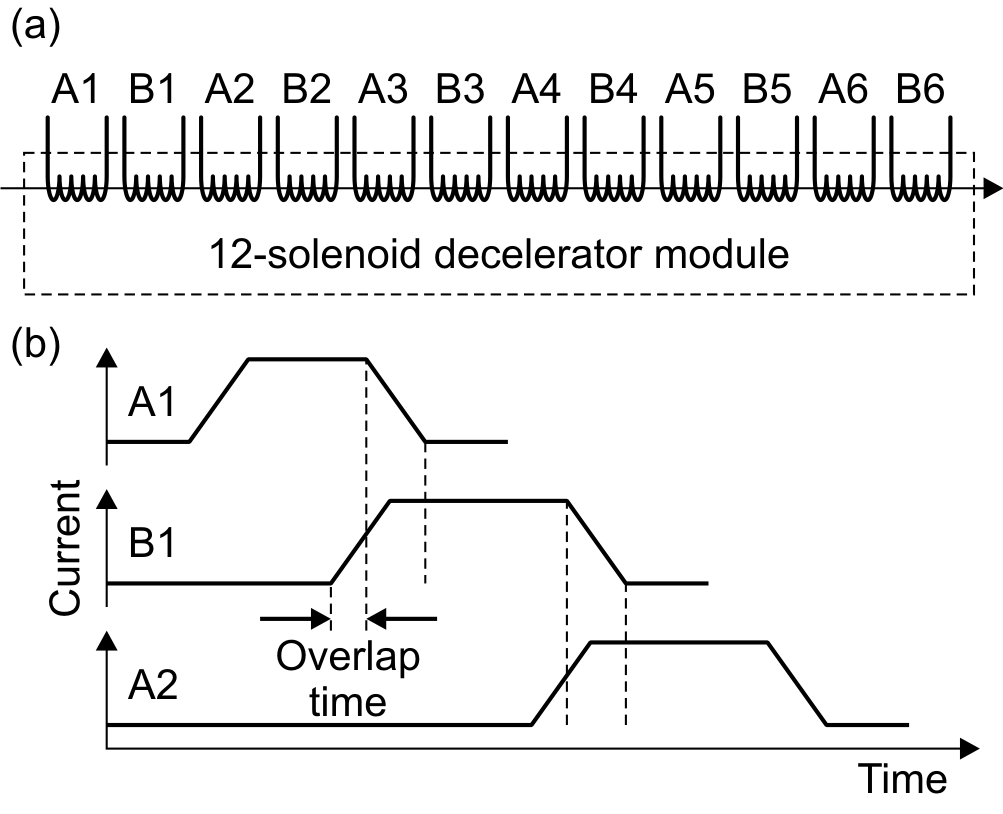}
\caption{\label{fig:electronics_overlap}
(a)~Alternating connection of the deceleration solenoids in one deceleration module to the power electronics drivers A and B.
(b)~To maintain a quantization field at all times during deceleration, the current through solenoid N+1 is switched on before switching off the current through solenoid N.
}
\end{figure}

To operate the multistage Zeeman decelerator in an efficient way, the particles must be prevented from moving over a long distance during the switch-off time of the field~\cite{wiederkehr10b}. Therefore, the decelerating magnetic field of each stage must be switched off as rapidly and completely as possible.

In \reffig{fig:electronics_scheme}, we show a diagram describing the operation principle of the electronic circuit used to achieve the sub-\unit{10}{\micro\second} rise and fall times of the current pulses applied to the deceleration solenoids. The circuit relies on a high-voltage supply for fast switching together with a low-voltage supply to maintain a constant current during the hold phase. To control the currents, insulated-gate bipolar transistor (IGBT) switches with integrated free-wheeling diodes (Semikron, SKM200GB176D) are used, which permit operation with pulsed currents of up to \unit{300}{\ampere}.

To rapidly ramp up the current through the solenoid, the IGBT switches K and S are closed, and the voltage $\mathrm{U}_\mathrm{K}$ is applied to the solenoid. The current through the solenoid increases as $\mathrm{I}_\mathrm{sol} = \mathrm{I}_0\,({1-\exp(-t/\tau)}) \approx \mathrm{I}_0\, t/\tau$ ($\mathrm{I}_0 = \mathrm{U}_\mathrm{K}/\mathrm{R}$, $\tau = \mathrm{L/R}$, $\mathrm{L}=\unit{21}{\micro\henry}$, $\mathrm{R}=\unit{0.24}{\ohm}$), the linear approximation being valid at short times. The current through the solenoid is monitored by the voltage drop across the shunt resistor $\mathrm{R_S}$ and fed to a comparator. When a preset value of the current has been reached at the end of the switch-on phase (1) (blue color code in \reffig{fig:electronics_scheme}), switch K is opened. The current commutes to branch (2) (green color code in \reffig{fig:electronics_scheme}) and is held constant because of the high energy stored in the capacitor $\mathrm{C_H}$.

To switch off the magnetic field produced by the solenoid, switch S is opened at the end of the hold phase (2). The decay of the magnetic field in the solenoid is accompanied by the build-up of an induction voltage which rises to a potential $\mathrm{U}_\mathrm{K}+\mathrm{U}_\mathrm{D}$ ($\mathrm{U}_\mathrm{D}$: forward voltage drop of diode $\mathrm{D_D}$) and opens a path (3) (red color code in \reffig{fig:electronics_scheme}) for the solenoid current into the storage capacitor $\mathrm{C_K}$. The current during the switch-off phase decreases with a time-reversed temporal behavior compared to the switch-on phase, and stops when all of the energy stored in the solenoid has been either transferred into capacitor $\mathrm{C_K}$ or dissipated as heat in the circuit.

The temporal magnetic-field profile of a deceleration solenoid as measured with a pickup coil is shown in \reffig{fig:fieldmeas}. At a kick voltage of $\mathrm{U_K}$=\unit{800}{\volt}, a rise and fall time of the \unit{2.2}{\tesla} magnetic field of \unit{8.5}{\mus} is deduced from this measurement. For typical durations of the hold phase of around \unit{30}{\mus}, the magnetic field strength remains constant, with a negligible decrease ($\leq 2\,\%$) resulting from a slow discharge of the hold capacitor $\mathrm{C_H}$. The undershoot at the end of the switch-off phase is attributed to the properties of diode $\mathrm{D_D}$ and has to be compensated to maintain a well-defined quantization field throughout the decelerator. This is achieved by applying a current to the adjacent deceleration solenoid. The weak oscillations following the undershoot, however, are an artifact caused by capacitive coupling to the pick-up coil.

The switching electronics are assembled into units capable of driving one deceleration module with 12 solenoids. Each of these electronics units consists of two drivers labeled A and B. Both drivers have their own, independent power electronics schematically drawn in \reffig{fig:electronics_module}~(a) and a control and interface circuitry as shown in \reffig{fig:electronics_module}~(b). This subdivision into two drivers A and B results from the necessity to maintain a quantization field at the position of the decelerated particles at all times, which is achieved by temporally overlapping the current pulses applied to adjacent solenoids as illustrated in \reffig{fig:electronics_overlap}. To minimize the total number of IGBT switches, a single IGBT switch K is used to pulse current through six of the deceleration solenoids in one module.

The individual electronics units are connected to a general-purpose pattern generator (Spincore PulseBlaster PB24-100-512) via a bus system, allowing for flexibility in the generation of deceleration pulse sequences. One electronics unit contains two identical interface and control circuits, each of them servicing one of the two drivers A and B. The bus interface decodes the signals from the pattern generator, and extracts the addresses of the electronics unit and the solenoid to be energized. To control the temporal current profile applied to the solenoid, the voltage $\mathrm{U_S}$ across the shunt resistor $\mathrm{R_S}$ is monitored, and IGBT switch K is opened if the preset value of the current (usually \unit{300}{\ampere}) is reached.

In addition to controlling and switching the power electronics, the control circuitry provides protection and interlock functions. The protection circuit inhibits gate drive to the IGBT if the repetition rate is too high, if current pulses longer than a specified maximal duration are applied, or if the interval between successive pulse commands is too short. In addition, the status output of the integrated monitor electronics of the IGBT driver is processed, thus inhibiting gate drive to the IGBT in case of an overcurrent condition. For monitoring purposes, the state of the electronics unit is displayed using status indicator LEDs, and the solenoid current measured across the shunt resistors is supplied to an analog interface.

To pulse current through the tower solenoids, a similar power electronics and control circuitry is used. However, the higher inductance (\unit{50}{\micro\henry} per solenoid) of these solenoids imposes longer switching times and necessitates the use of larger storage capacitors to provide a sufficiently high energy reservoir. Therefore, a single-channel electronics unit supplies current to the two solenoids of each tower which are connected in series.

\section{Results and Discussion}
\label{sec:results}

The deceleration process was characterized by a combination of time-of-flight and imaging techniques. To generate a deceleration pulse sequence, the synchronous particle with a given initial velocity was propagated along the axis of the decelerator in a one-dimensional trajectory simulation taking into account the finite rise and fall times of the decelerating fields. As explained in~\cite{wiederkehr10b}, we define the phase angle $\phi_0$ in a deceleration sequence as the reduced position of the synchronous particle at the time the switch-off of the field is initiated. When a measurement is referred to as having been performed at a phase angle $\phi_0$, this means that all deceleration stages throughout the decelerator were operated with the same phase angle $\phi_0$. Parameters which were systematically varied in the measurements include the phase angle and the initial velocity of the synchronous particle.

\subsection{Time-of-flight distribution of decelerated atoms}
\label{sec:results:basics}

\begin{figure}
\includegraphics{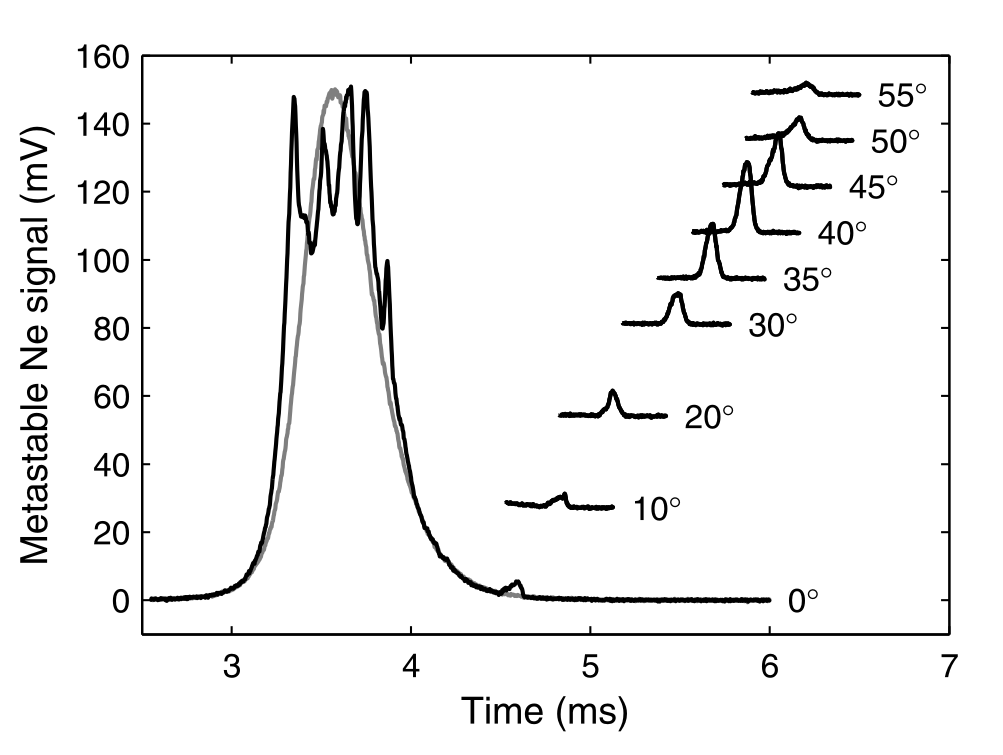}
\caption{\label{fig:tof_overview}
Time-of-flight distribution of metastable neon atoms recorded with the MCP detector placed \unit{370}{\mm} downstream of the decelerator exit. The gray trace was measured with the multistage decelerator inactive. For the other traces, the deceleration sequence was calculated for an initial velocity of the synchronous particle of \unit{580}{\mps}, and the phase angle was varied between \unit{0}{\degree}--\unit{55}{\degree} as indicated next to the trace. For clarity, the traces recorded with phase angle \unit{10}{\degree}--\unit{55}{\degree} are vertically offset, and only a time window centered at the arrival time of the decelerated atoms is shown.
}
\end{figure}

Time-of-flight distributions recorded for different settings of the decelerator are compared in \reffig{fig:tof_overview} to a time-of-flight distribution obtained with the decelerator inactive (gray trace). In these measurements, time zero corresponds to the trigger of the pulsed valve. For a supersonic beam at high Mach number, the velocity distribution can be approximated by a Gaussian distribution.

When the decelerator is turned on, the appearance of the main time-of-flight peak changes qualitatively (see lowest black trace in \reffig{fig:tof_overview} recorded at a phase angle of \unit{0}{\degree}). The rich substructure of this peak is caused by transverse focusing and guiding of the metastable atoms by the pulsed magnetic fields~\cite{hogan07a, hogan08c}. In this time-of-flight profile, an additional peak is observed at a later time (\unit{\approx 4.6}{\ms}), corresponding to neon atoms that are decelerated in a phase-stable manner at $\phi_0=\unit{0}{\degree}$. It is, however, not possible to directly extract the velocity of the decelerated atoms from their arrival time with respect to the time of flight of the undecelerated atoms, because the change in energy and velocity per time interval is not constant during the propagation of the atoms through the decelerator. Nevertheless, when all deceleration stages are operated at the same phase angle, a later arrival time of the decelerated packet corresponds to a lower final velocity. Comparing the signal of the decelerated-atom peaks in \reffig{fig:tof_overview} to the signal of atoms with the decelerator inactive, it should be noted that all five magnetic sublevels of the \tp{2} state as well as the \tp{0} state contribute to the latter, whereas only atoms in the state \tp{2}, $M_J=2$ contribute to the peak of decelerated atoms.

\begin{figure}
\includegraphics{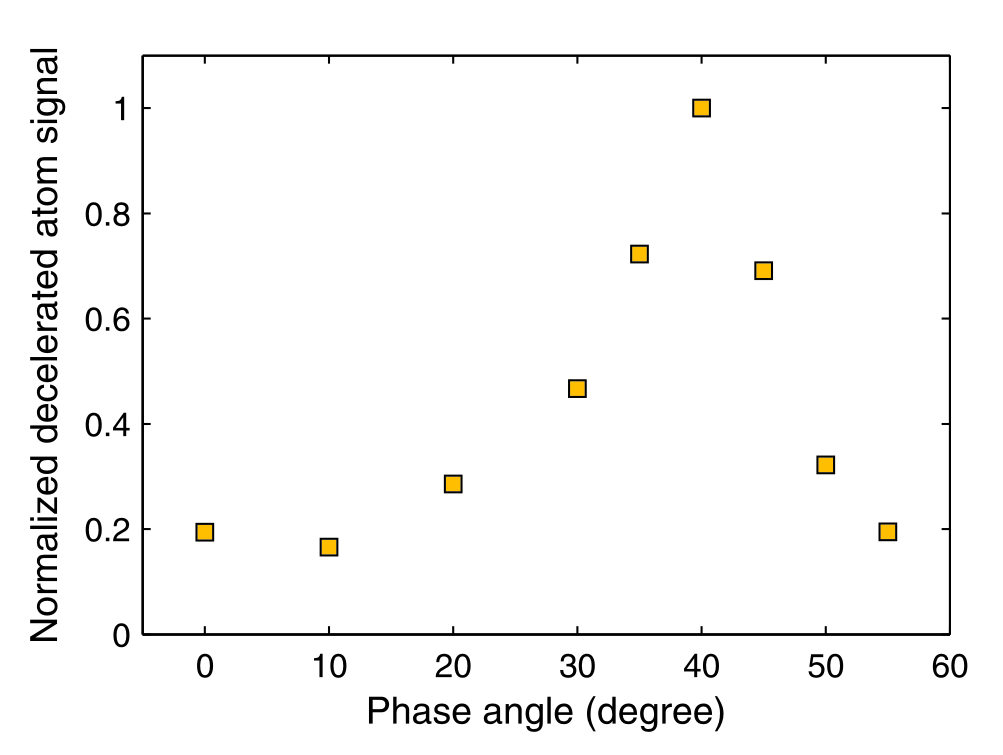}
\caption{\label{fig:decel_result_number}
Intensity of the decelerated atom signal determined from the area of the decelerated-atom peaks in the time-of-flight distributions. The deceleration sequences were generated for an initial velocity of the synchronous particle of \unit{580}{\mps}.
\newline
}
\end{figure}

When a deceleration sequence with an increased phase angle is applied, a larger amount of kinetic energy is removed per deceleration stage, and the particles within the phase-stable region around the synchronous particle are decelerated to lower velocities. This behavior can be observed in the time-of-flight distributions shown in \reffig{fig:tof_overview}, where the arrival time of the decelerated atoms increases with increasing phase angle. At the same time, a one-dimensional model of a multistage Zeeman decelerator predicts a reduction of the phase-stable volume accepted by the decelerator when the phase angle is increased~\cite{wiederkehr10b}; one might thus expect the largest signal of decelerated atoms at \unit{0}{\degree} phase angle. The intensity of the decelerated-atom peak shown in \reffig{fig:decel_result_number}, however, does not show such a monotonic behavior. Instead, we find that the decelerator appears to operate most efficiently at a phase angle of \unit{40}{\degree}. A careful analysis of the three-dimensional particle dynamics in the multistage Zeeman decelerator reveals strong transverse effects not accounted for in the one-dimensional model. While transverse focusing and defocusing effects reduce the phase-stable volume at low phase angle, the finite rise and fall times of the pulsed magnetic fields in the deceleration solenoids tend to compensate this to some degree and increase the phase-stable volume. Altogether, three-dimensional trajectory simulations demonstrate that with our solenoid configuration the multistage Zeeman decelerator has the largest accepted phase-stable volume ($\approx$\unit{10}{\mm^3}$\times$\unit{200}{(\mps)^3}) for phase angles in the range \unit{30}{\degree}--\unit{45}{\degree}~\cite{wiederkehr10b}, in agreement with our experimental findings (see Sections~\ref{sec:results:longvel} and \ref{sec:results:imaging}). The present analysis further confirms that the results of the phase-stability analysis performed for D~$(1^2S_{1/2})$ atoms can be transferred to the deceleration of heavier species in significantly longer decelerators.

At the base pressure of the decelerator of \unit{2\times10^{-7}}{\milli\bbar}, losses by collisions with the background gas are negligible. Experiments carried out at higher pressure by introducing nitrogen into the vacuum chamber through a needle valve indicate that collisional losses become significant at pressures exceeding \unit{2\times10^{-6}}{\milli\bbar}.

\subsection{Determination of the longitudinal velocity distribution}
\label{sec:results:longvel}

The direct detection of metastable neon atoms impinging on the MCP offers a simple way to experimentally characterize the velocity distribution of the beam: the velocity distribution is mapped onto an arrival-time distribution by free propagation. By recording time-of-flight distributions for several distances of free propagation (\unit{200}{\mm}, \unit{370}{\mm}, and \unit{505}{\mm}) between the exit of the decelerator and the MCP detector, it is possible to reconstruct the velocity distribution.

\begin{figure}
\includegraphics{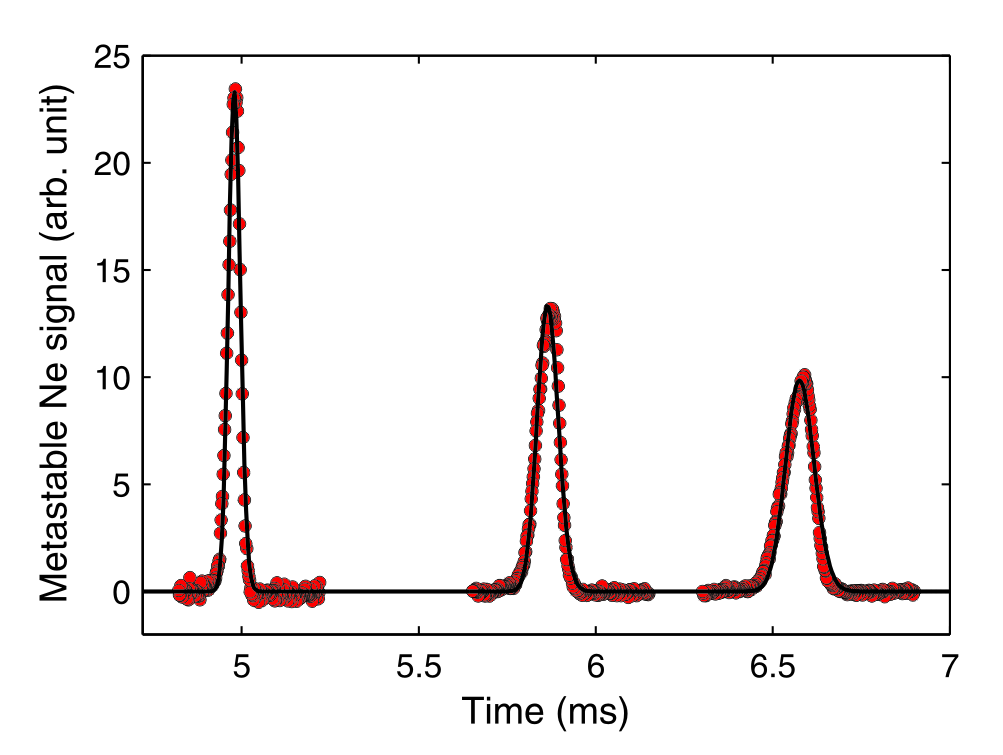}
\caption{\label{fig:tof_fitmodel}
Time-of-flight distributions of metastable neon atoms recorded with \unit{200}{\mm}, \unit{370}{\mm}, and \unit{505}{\mm}  of free propagation from the exit of the decelerator to the MCP detector. The deceleration pulse sequence was generated for an initial velocity of \unit{580}{\mps} and a phase angle of \unit{40}{\degree}. For clarity, only a portion of the time-of-flight distribution centered at the arrival time of the decelerated atoms is shown. The solid curve is a fit of $\sum_i S(L_i,t)\,\dd{t}$, with $S(L,t)\,\dd{t}$ defined by \refeq{eq:tof_convoluted}, to these data.
}
\end{figure}

The time-of-flight profiles of decelerated neon atoms recorded at a phase angle of \unit{40}{\degree} for these three different free-propagation distances are compared in \reffig{fig:tof_fitmodel}. Similar profiles were recorded for several phase angles beween \unit{0}{\degree} and \unit{55}{\degree}. The increasing temporal spread of the packet of decelerated atoms during its propagation to the detector is caused by velocity dispersion. To extract the longitudinal temperature from the time-of-flight profiles, two methods were used. The first method, based on Gaussian fits to the observed time-of-flight distributions, offers the advantage of being robust and simple, and enables rapid diagnostics of the deceleration process during data acquisition. The second, more accurate method, relies on a complete model of the particle propagation including spatial and velocity distributions. For the analysis, all traces were individually normalized to the signal of decelerated atoms to account for small variations in the operation conditions of the metastable neon source.

\subsubsection{Velocity of decelerated atoms}

\begin{figure}
\includegraphics{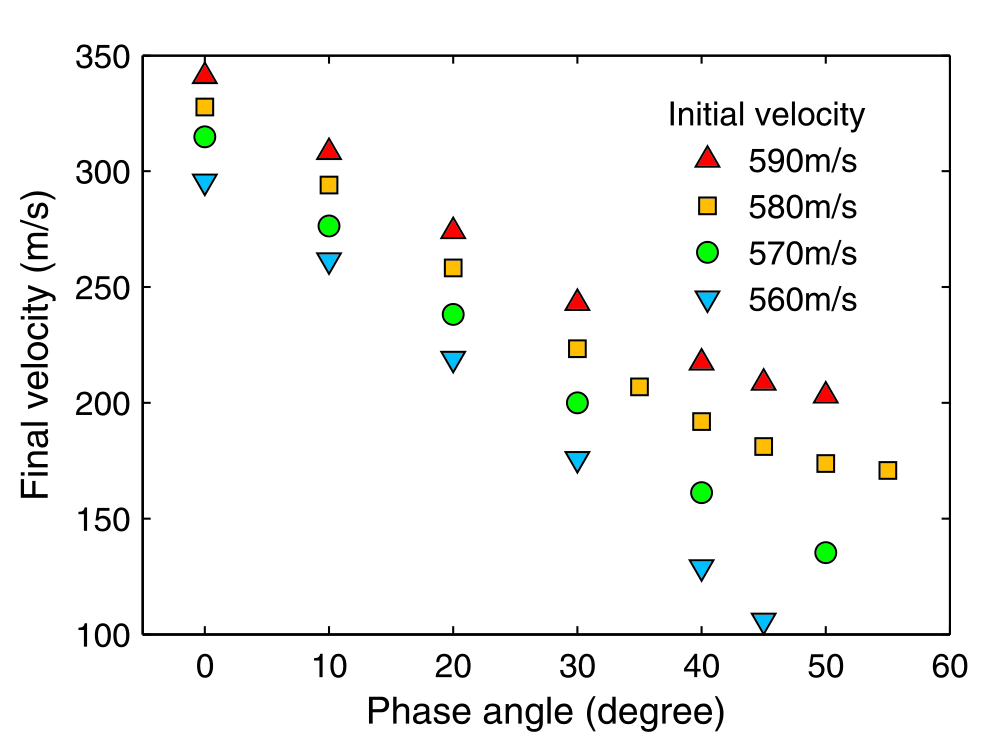}
\caption{\label{fig:decel_result_velocity}
Measured final velocities of decelerated neon atoms determined from their arrival time at the detector with different distances of free propagation. In the case of the lowest final velocity, \unit{106}{\mps}, more than 95\,\% of the initial kinetic energy of the beam was removed in the 91-stage decelerator.
}
\end{figure}

For a preliminary analysis, a fit of Gaussian functions
\begin{equation}
\label{eq:tof_simple}
S(t)\,\dd{t} = \frac{1}{\sqrt{2\pi}\sigma_t}\,\exp\left(-\frac{(t-t_c)^2}{2\sigma_t^2}\right)\,\dd{t}
\end{equation}
was made to the individual peaks of decelerated atoms in the time-of-flight signal recorded for different free-propagation distances. The mean velocity $\bar{v}$ and the velocity spread $\sigma_v$ were then extracted from an analysis of the peak position $t_c(L)$ and the width $\sigma_t(L)$ of the decelerated-atom peak in the time-of-flight distribution for each free-propagation distance $L$. The mean velocity of the decelerated atoms was determined by fitting
\begin{equation}
\label{eq:tof_simple_tc}
t_c(L) = t_0 + \frac{1}{\bar{v}}\,L
\end{equation}
to the data, where $t_0$ is the time at which the free propagation starts (i.e., the time the decelerated atoms have reached the exit of the decelerator). The velocity spread of the decelerated atoms was then obtained by fitting
\begin{equation}
\label{eq:tof_simple_sigmat}
\sigma_t(L)^2 = \frac{\sigma_v^4}{\bar{v}^4} \, L^2 + \frac{\sigma_z^2}{\bar{v}^2}
\end{equation}
to the data, where $\sigma_z$ is the width of the Gaussian packet at the exit of the decelerator.
In \reffig{fig:decel_result_velocity}, the velocities of the decelerated atoms for different phase angles are compared. These velocities were obtained by analysing measuremets for three different free-propagation distances and correspond to the final velocities for which the deceleration pulse sequences were optimized to within \unit{1}{\mps}. This indicates that the multistage decelerator and the deceleration process are very well characterized and atom bunches of a desired final velocity can be reliably generated. The analysis also confirms that a higher phase angle results in a lower velocity of the decelerated atoms, as expected from simple considerations.

\subsubsection{Modeling the time-of-flight distribution}
To model the arrival-time distribution at the detector, the longitudinal velocity distribution of the packet of atoms decelerated to a velocity $\bar{v}$ is described by a Gaussian distribution
\begin{equation}
f(v)\,\dd{v} = \frac{1}{\sqrt{2\pi}\sigma_v}\, \exp\left(-\frac{(v-\bar{v})^2}{2\sigma_v^2}\right)\,\dd{v},
\end{equation}
where $\sigma_v^2$ is the variance in the final longitudinal velocity. Free propagation over a distance $L$ maps the velocity distribution $f(v)\,\dd{v}$ to an arrival-time distribution at the detector,
\begin{equation}
S(L,t)\,\dd{t} = f(L\,t^{-1}) \, Lt^{-2}\,\dd{t}.
\end{equation}
The longitudinal extent of the cloud of decelerated atoms at the beginning of the free-propagation phase can be approximated by a Gaussian distribution,
\begin{equation}
g(z)\,\dd{z} = \frac{1}{\sqrt{2\pi}\sigma_z}\, \exp\left(-\frac{z^2}{2\sigma_z^2}\right)\,\dd{z}.
\end{equation}
The arrival-time distribution is obtained by the convolution
\begin{equation}
S(L,t)\,\dd{t} = \int\limits_{-\infty}^\infty \left[f(\frac{L-z}{t})\,Lt^{-2}\,g(z)\right]\,\dd{z}\, \dd{t},
\end{equation}
which, in the case of two Gaussian distributions, yields
\begin{multline}
\label{eq:tof_convoluted}
S(L,t)\,\dd{t} = \\
\frac{1}{\sqrt{2\pi}}
\frac{L \, t \, \sigma_v^2 + \bar{v}\,\sigma_z^2}{(t^2 \, \sigma_v^2 + \sigma_z^2)^{3/2}}\,
\exp\left(-\frac{(L-\bar{v}\,t)^2}{2(t^2\,\sigma_v^2+\sigma_z^2)}\right)\,\dd{t}.
\end{multline}
This expression has the same form as \refeq{eq:tof_simple}, which justifies the use of \refeq{eq:tof_simple_tc} and \refeq{eq:tof_simple_sigmat} in the first analysis, but links the observed time-of-flight profiles to both the spatial and velocity distributions of the decelerated atoms.

\subsubsection{Velocity distributions from time-of-flight measurements}

\begin{figure*}
\includegraphics{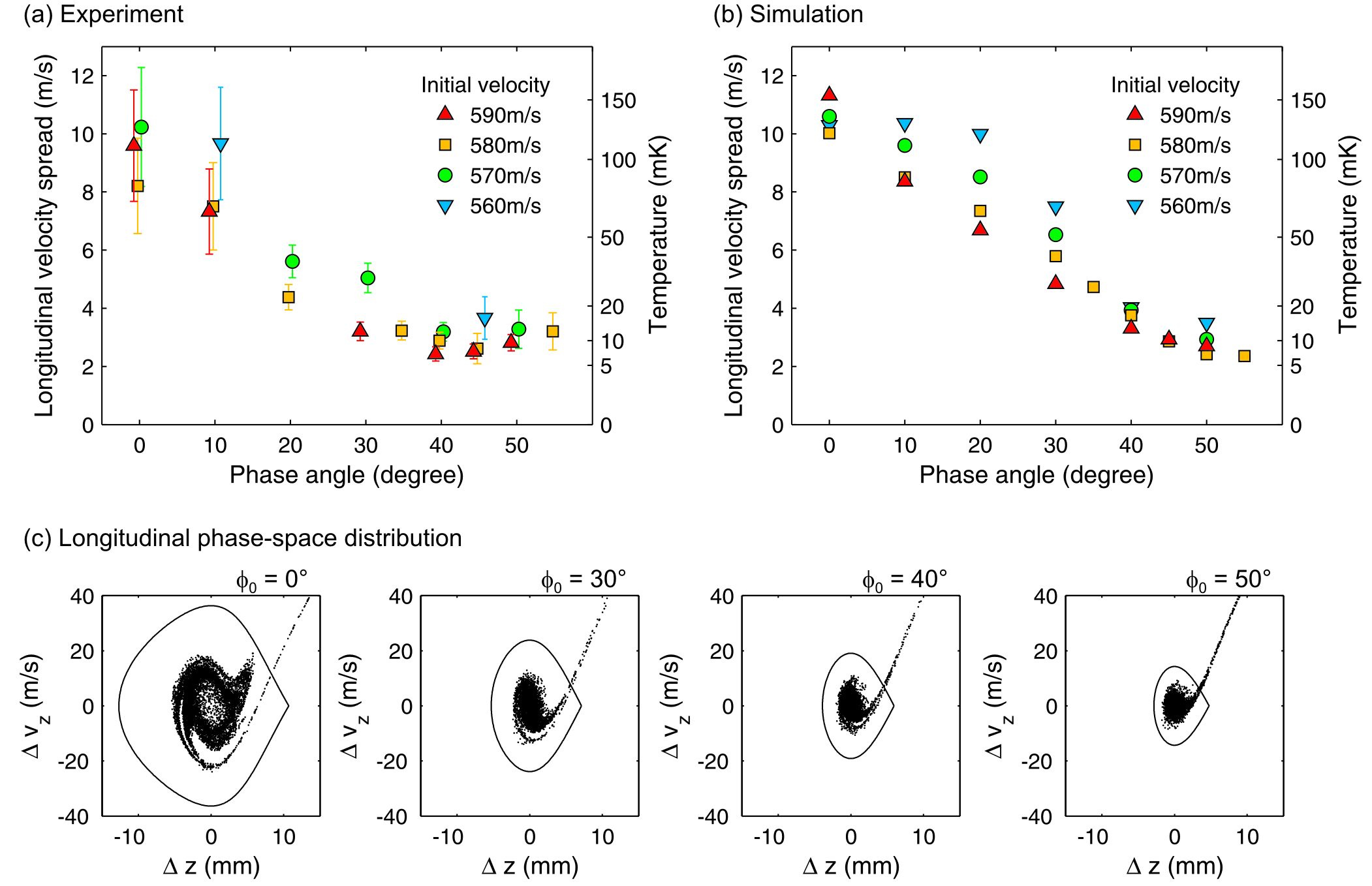}
\caption{\label{fig:tof_spread}
(a)~Velocity spread of the decelerated atoms determined by fitting the model function $\sum_i S(L_i,t)\,\dd{t}$ to the time-of-flight distributions. The data points in a group of measurements at the same phase angle but different initial velocities are slightly displaced horizontally from each other for clarity. The error bars estimate the accuracy of the fit in reproducing the data. The temperature is determined from the velocity spread using the relation  $k_\mathrm{B}T=\frac{1}{2}m\sigma_v^2$.
(b)~Velocity spread and (c)~phase-space distribution of the decelerated atoms obtained from three-dimensional numerical particle-trajectory simulations.
}
\end{figure*}

To reconstruct the velocity distribution of the decelerated atoms, $\sum_{i} S(L_i,t)\,\dd{t}$ $(i=1,2,3)$ was fitted to the time-of-flight distributions measured with free-propagation distances $L_1=\unit{200}{\mm}$, $L_2=\unit{370}{\mm}$, and $L_3=\unit{505}{\mm}$ as shown in \reffig{fig:tof_fitmodel}. This procedure is more sensitive to changes of the peak shape with propagation than fitting individual Gaussian functions to the peaks in the time-of-flight traces. Thus, a good agreement between the fit and the experimental data, as shown in \reffig{fig:tof_fitmodel}, is only obtained if the model functions used for the velocity distribution and the initial spatial distribution are good approximations to the real phase-space distribution of the decelerated atoms. The parameters which can be obtained from the fit are the velocity of the decelerated atoms $v_\mathrm{decel}$ and their velocity spread $\sigma_v$. Although the model function \refeq{eq:tof_convoluted} depends on the longitudinal position spread $\sigma_z$ at the exit of the decelerator, the fits are rather insensitive to this parameter because the temporal spread of the packet in the time-of-flight distribution is dominated by velocity dispersion, even for the shortest distance of \unit{200}{\mm}. Therefore the longitudinal spread at the exit of the decelerator obtained from trajectory simulations, which was found to decrease from $\sigma_z=\unit{2.5}{\mm}$ at $\phi_0=\unit{0}{\degree}$ to $\sigma_z=\unit{1}{\mm}$ at $\phi_0=\unit{50}{\degree}$, was used as a fixed input parameter to the fit.

In \reffig{fig:tof_spread}, the experimentally determined dependence of the velocity spread of the decelerated atoms on the phase angle is compared to the outcome of three-dimensional numerical particle-trajectory simulations. The good overall agreement indicates that the simulations reproduce the main features of the deceleration process adequately. For phase angles in the range of \unit{30}{\degree}--\unit{50}{\degree}, where the decelerator is operated most efficiently (see \reffig{fig:decel_result_number} and Wiederkehr et al.~\cite{wiederkehr10b}), the longitudinal temperature $T$ of the decelerated atoms determined from $k_\mathrm{B}T = \frac{1}{2}m\sigma_v^2$ is found to be only \unit{10}{\milli\kelvin}. This is less than typical depths of magnetic traps of around \unit{100}{\milli\kelvin}~\cite{sawyer08a, hogan08d, wiederkehr10a}, so that efficient trap loading following the procedure described in~\cite{hogan08d, wiederkehr10a} can be envisaged.

The temperature of the decelerated atoms is found to decrease with increasing phase angle. Following a simple model of the Zeeman decelerator, which assumes infinitely short rise and fall times of the decelerating fields, one expects the final temperature to directly reflect the phase-space acceptance of the decelerator. Hence, at a fixed phase angle, it should be independent of the initial velocity of the atoms. The simulations, however, indicate an influence of the initial velocity of the synchronous particle, for which the deceleration pulse sequence is calculated, on the temperature of the decelerated atoms. This effect is caused by the finite rise and fall times of the decelerating magnetic field. Atoms which move at a higher velocity propagate further into the solenoid during the switch-off time of the field. Thus, they experience a larger effective phase angle, which reduces the phase-space acceptance of the decelerator, and results in a lower temperature.

For comparison to previous work~\cite{wiederkehr10b}, the longitudinal phase-space distribution of decelerated atoms obtained from trajectory simulations and the separatrix obtained from a one-dimensional phase-stability analysis are presented in \reffig{fig:tof_spread}~(c). The phase-stable area in this diagram is found to be smaller than the one-dimensional separatrix at each phase angle plotted. At low phase angles (here shown for \unit{0}{\degree}), a reduction of the number of particles near the phase-space position of the synchronous particle, i.e., in the central region of the separatrix, is observed. As explained in~\cite{wiederkehr10b}, these differences between the predictions of the one-dimensional model and the phase-stable volume of a \enquote{real} multistage Zeeman decelerator are caused by an interplay between the three-dimensional magnetic field distribution and the finite rise and fall time of the field. At low phase angles, the convex, defocussing nature of the magnetic-field distribution at large distances from the center of the solenoids cannot be compensated any more and a loss of particles results. The interruption of the otherwise regular array of deceleration solenoids by the towers can cause an additional reduction of the phase-stable volume. Alltogether, the analysis demonstrates that the results of the extensive study of phase-stability performed for D atoms~\cite{wiederkehr10b} can be transferred to the description of a modular multistage Zeeman decelerator for other species.

\subsection{Imaging decelerated neon atoms}
\label{sec:results:imaging}

The determination of the longitudinal velocity distribution of decelerated atoms by the method discussed in \refsec{sec:results:longvel} was greatly faciliated by the direct detection of the metastable neon atoms impinging on the active surface of the MCP detector. The time-of-flight measurements of the longitudinal velocity distribution can be extended in a straightforward way to also determine the transverse spatial and velocity distributions of the decelerated atoms. For this purpose, a MCP equipped with a phosphor screen behind the active layers was installed, and the phosphor screen was imaged from behind with a CCD camera.

\begin{figure}
\includegraphics{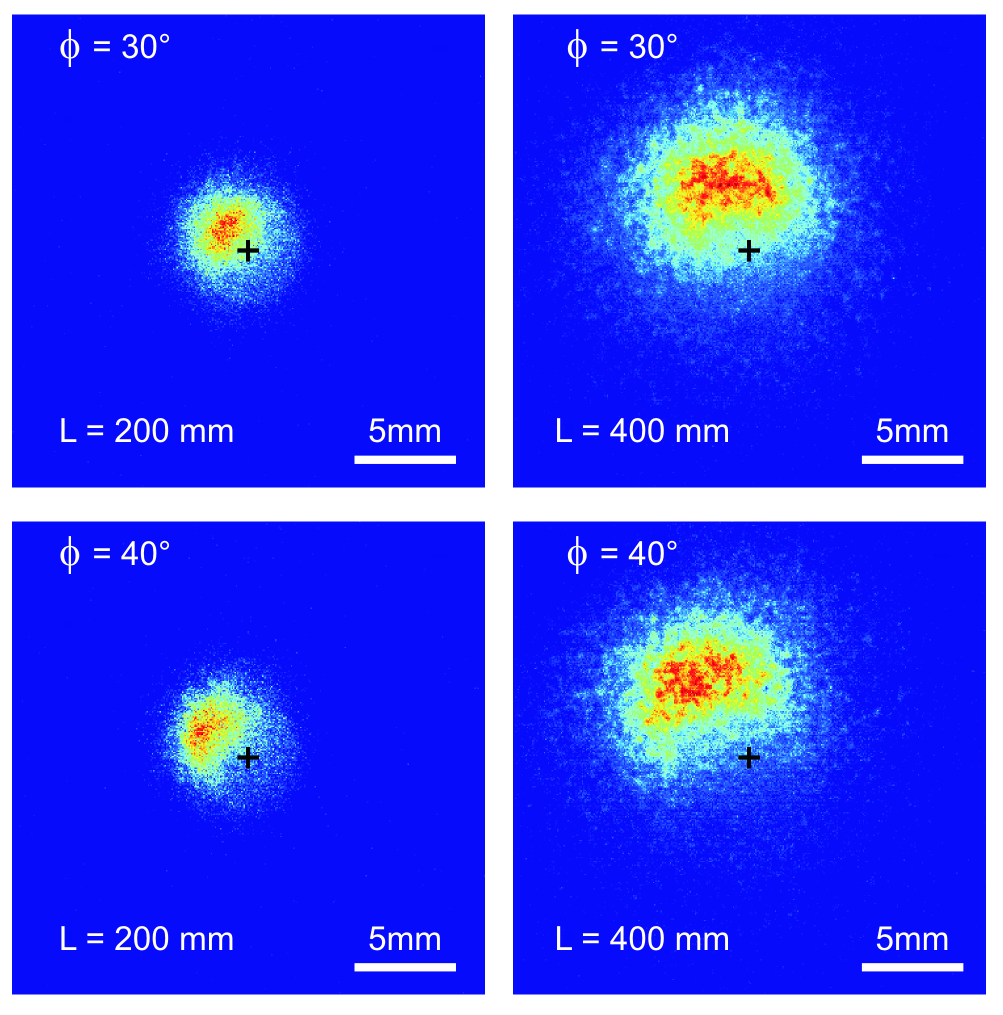}
\caption{\label{fig:imaging}
Background-subtracted false-color images of metastable neon atoms decelerated to final velocities of \unit{220}{\mps} and \unit{180}{\mps} at a phase angle of \unit{30}{\degree} and \unit{40}{\degree}, respectively. L is the distance of free propagation between the exit of the decelerator and the MCP. The camera exposure time of \unit{100-200}{\mus} was matched to the arrival-time spread of the decelerated atoms. The black cross indicates the point of intersection of the decelerator axis and the MCP detector surface.
}
\end{figure}

In \reffig{fig:imaging}, we show background-subtracted false-color images of decelerated metastable neon atoms obtained by this method. Such images were recorded for deceleration pulse sequences with a phase angle between \unit{0}{\degree} and \unit{50}{\degree}, and for free-propagation distances of \unit{200}{\mm} and \unit{400}{\mm} from the exit of the decelerator to the MCP. These images serve the purpose of characterizing the beam temperature and aid in the alignment of the decelerator.

In the images shown in \reffig{fig:imaging}, a displacement of the decelerated atom cloud from the reference position, given by the point of intersection of the decelerator axis and the MCP detector surface, is apparent. The shape of the atom cloud is also slightly distorted, the effect being most pronounced at small phase angles. Both effects were found to sensitively depend on the transverse alignment and tilting of the deceleration modules with respect to each other, and the images shown in \reffig{fig:imaging} represent the best compromise achieved when optimizing the alignment for the most relevant phase-angle range of between \unit{30}{\degree} and \unit{40}{\degree}. The asymmetry and displacement is likely caused by the limited precision in the positioning of the current-carrying solenoids. Since these effects become visible only at long distances from the exit of the multistage Zeeman decelerator, we do not expect them to hamper the use of the decelerator in high-resolution spectroscopy, in collision experiments, or in trapping experiments, which would ideally be performed at shorter distances.

To obtain the transverse temperature and the transverse position spread of the packet of decelerated atoms at the exit of the decelerator, the transverse extension of the slow atom cloud was studied as a function of propagation distance. The analysis was performed for phase angles in the range \unit{30}{\degree}--\unit{45}{\degree} in which the decelerator is operated most efficiently. Assuming a Gaussian distribution, a transverse temperature of \unit{5-10}{\milli\kelvin} can be determined in combination with a transverse spread at the exit of the decelerator of about \unit{1}{\mm}. These values are in agreement with the predictions of three-dimensional particle-trajectory simulations. This agreement demonstrates that the phase-space distribution of the decelerated atoms can be completely reconstructed from a combination of time-of-flight and imaging measurements.

\subsection{Isotope-selective multistage Zeeman deceleration of metastable neon}
\label{sec:results:isotopes}

\begin{figure}
\includegraphics{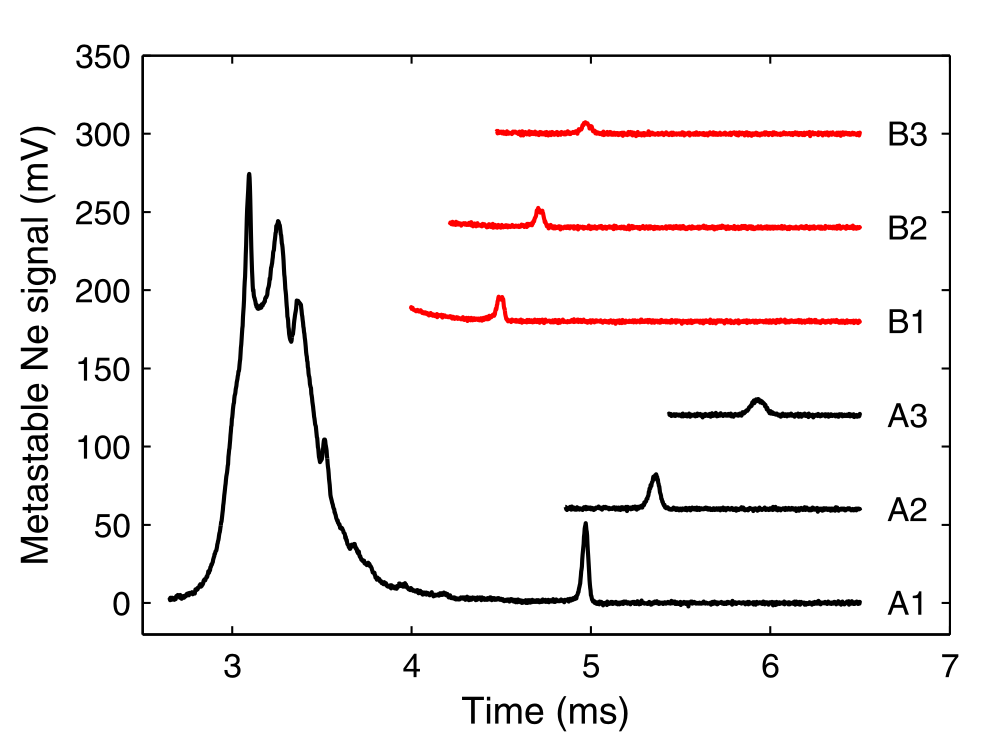}
\caption{\label{fig:isotopes_tof}
Time-of-flight distribution of metastable neon atoms. The lower traces (A1--A3) were recorded with deceleration pulse sequences generated for \nei{20},  whereas the upper traces (B1--B3) were recorded with pulse sequences for \nei{22}. The initial velocity of the synchronous particle was \unit{580}{\mps} (A1, B1), \unit{570}{\mps} (A2, B2), and \unit{560}{\mps} (A3, B3), and a constant phase angle of \unit{40}{\degree} was used. The final velocities of the decelerated atoms are \unit{191}{\mps} (A1), \unit{161}{\mps} (A2), \unit{129}{\mps} (A3), \unit{248}{\mps} (B1), \unit{226}{\mps} (B2), and \unit{202}{\mps} (B3). For clarity, the traces recorded with different deceleration pulse sequences are vertically offset.
}
\end{figure}

\begin{figure*}
\includegraphics{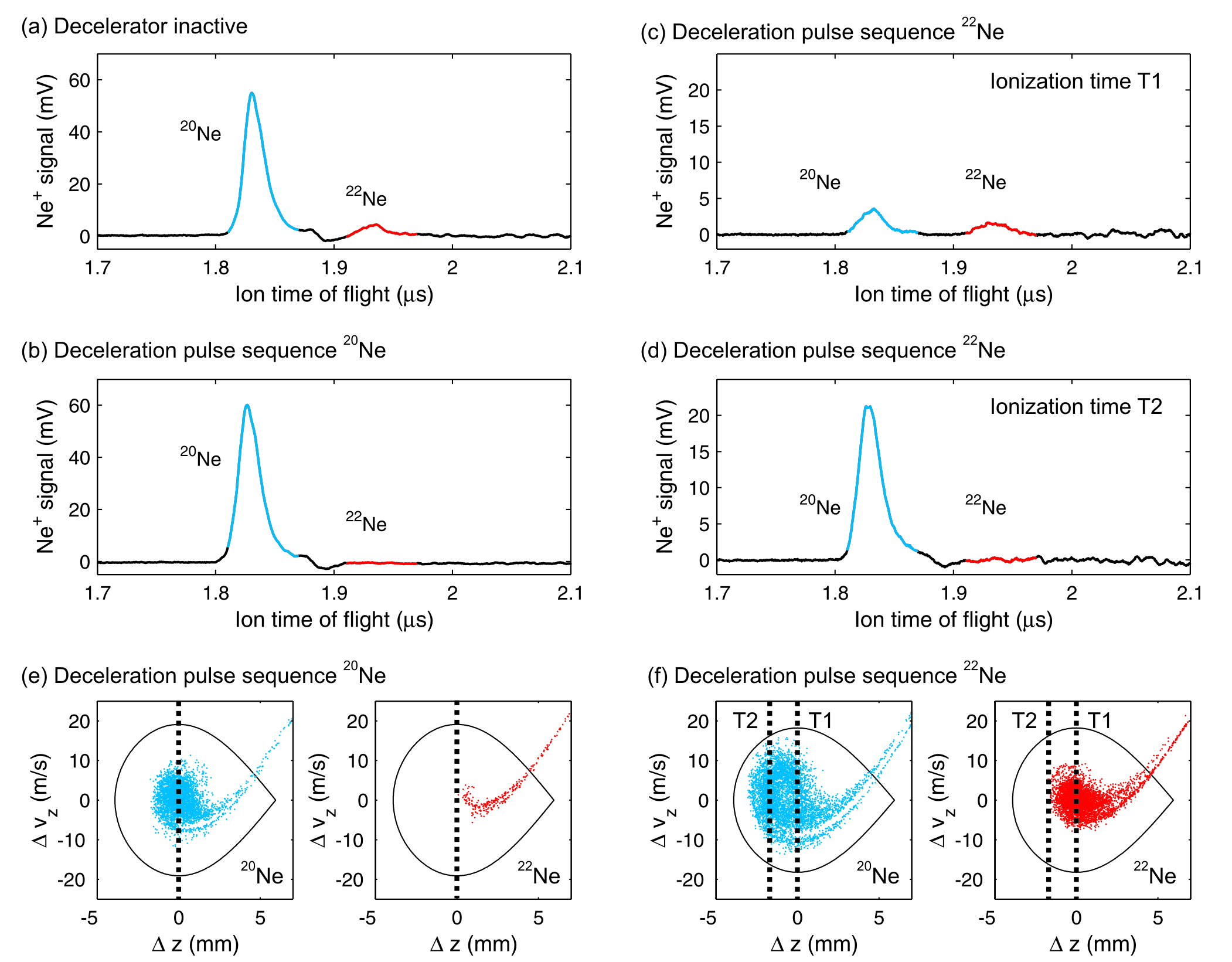}
\caption{\label{fig:isotopes_ions}
(a)--(d)~Time-of-flight mass-spectrometer signal obtained by photoionizing metastable neon atoms in the beam.
The deceleration pulse sequences used were generated for a phase angle of \unit{40}{\degree} and an initial velocity of the synchronous particle of \unit{580}{\mps}.
(e)--(f)~Phase-space distributions of decelerated atoms obtained from particle-trajectory simulations, which were performed for an equal number of \nei{20} and \nei{22} atoms emerging from the pulsed valve. The dashed lines indicate the position in phase space overlapping with the ionization volume of the laser.
}
\end{figure*}

The significant natural abundance of two stable isotopes (\nei{20}: 90.5\,\%, \nei{22}: 9.3\,\%) of neon opens up the possibility to investigate whether multistage Zeeman deceleration can be used to separate species of different magnetic-moment-to-mass ratio. For this purpose, pulse sequences generated for a synchronous particle of each of the two isotopes were used, and the time-of-flight traces of decelerated metastable neon atoms were analyzed. Typical results are displayed in \reffig{fig:isotopes_tof}. Because the same amount of kinetic energy is removed from the synchronous particle per deceleration stage if the decelerator is operated at constant phase angle, a larger velocity change is expected for the lighter isotope. Comparing the time-of-flight distributions recorded for the same initial velocity but different mass of the synchronous particle (i.e., trace A1--A3 with B1--B3), the later arrival time of the decelerated atom peaks with the pulse sequences generated for \nei{20} indicates a lower final velocity. This is confirmed by measurements with different free-propagation distances (see caption of \reffig{fig:isotopes_tof} for the final velocities of the decelerated neon atoms determined experimentally).

However, these time-of-flight measurements alone do not allow one to disentangle the isotopic composition of the decelerated packet. To achieve this, a complementary measurement based on photoionization was used, which allowed isotopic separation of the resulting photoions in a time-of-flight mass spectrometer. The results are displayed in \reffig{fig:isotopes_ions}, and show the time-of-flight distribution of ions produced by pulsed-field ionization after photoexcitation of neon atoms from the \tp{2} state to Rydberg states of principal quantum number $n=24$.  The pulsed field also served the purpose of extracting the Ne$^+$ ions out of the metastable atom sample towards the MCP detector.

In a first measurement shown in \reffig{fig:isotopes_ions}~(a), the decelerator was inactive, and the delay of the laser pulse with respect to the valve opening time was set so as to photoexcite atoms at the maximum of the gas pulse (\unit{3.2}{\ms} in \reffig{fig:isotopes_tof}). In this measurement, the ratio of the two isotopes, as determined from the area of the corresponding peaks in the time-of-flight mass-spectrometer signal, was found to be in good agreement with that expected from their natural abundance.

In a second measurement, a deceleration pulse sequence generated for \nei{20} was applied, and the laser delay was set to detect atoms at the center of the decelerated atom peak (\unit{4.95}{\ms} in trace A1 in \reffig{fig:isotopes_tof}). In the ion time-of-flight distribution shown in \reffig{fig:isotopes_ions}~(b), only \nei{20} is observed, and the signal at the expected arrival time of \nei{22} is reduced to the background level.

In the last measurement, a deceleration pulse sequence generated for \nei{22} was applied. In this case, differences in the ion time-of-flight signal are observed depending on the time of photoexcitation within the decelerated atom peak in trace B1 in \reffig{fig:isotopes_tof}. In the ion time-of-flight signal shown in \reffig{fig:isotopes_ions}~(c), which was recorded with the laser set to an early time T1 of the decelerated atom pulse, both \nei{20} and \nei{22} are observed, but the fraction of \nei{22} (22\,\%) exceeds the natural abundance of 9.3\,\%. If the laser is set to a late time T2 of the decelerated atom pulse (see \reffig{fig:isotopes_ions}~(d)), however, only \nei{20} is observed in the time-of-flight trace.

The experimental results obtained with both deceleration pulse sequences can be understood from the phase-space diagrams depicted in panels (e) and (f) of \reffig{fig:isotopes_ions}, which were obtained from particle-trajectory simulations. Whereas efficient separation is possible with deceleration pulse sequences optimized for the lighter isotope, only partial separation can be achieved with the sequence optimized for the heavier isotope. This overall behavior shows similarities to the behavior observed by Bethlem et al.\ in the multistage Stark deceleration of ammonia isotopomers~\cite{bethlem02a}. For a given deceleration pulse sequence, the particle with the smaller magnetic-moment-to-mass ratio (in this case \nei{22}) must lose a larger amount of kinetic energy to stay synchronous with the switching of the fields. Therefore, it experiences a pulse sequence with a larger phase angle during propagation through the decelerator, which leads to a smaller phase-space aceptance. Generally speaking, lighter isotopes are more strongly influenced by the field gradients than the heavier ones and, unlike the heavier ones, can be decelerated even by sequences optimized for different isotopes. Separation of the lightest isotope by multistage Zeeman deceleration is thus possible using deceleration sequences optimized for their larger magnetic-moment-to-mass ratio.

\section{Conclusion and Outlook}
\label{sec:outlook}

In this work on multistage Zeeman deceleration of metastable neon atoms, the properties of the decelerated atoms were investigated using time-of-flight and imaging detection techniques. These methods enabled the complete characterization of the longitudinal and transverse velocity distributions of the decelerated atoms. With their widely tunable velocity and with longitudinal and transverse temperatures of about \unit{10}{\milli\kelvin}, cold atom and molecule samples produced by multistage Zeeman deceleration are expected to find applications in the study of reaction dynamics at low temperatures, in precision spectroscopy, and as starting point for further cooling steps. Multistage Zeeman deceleration leads to the production of samples with complete state selectivity and complete angular-momentum orientation as demonstrated here for the $M_J=2$ level of metastable neon. The potential of such samples for reaction dynamics is evident, but so far unexplored.

\begin{acknowledgments}
The authors thank K.~Dulitz (Oxford University) for stimulating discussions and comments on the manuscript. This work is supported by the Swiss National Science Foundation (Project No.\ 200020-132688) and the European Research Council advanced grant program (Project No.\ 228286). M.~M.\ thanks ETH Z\"urich for the support through an ETH fellowship.
\end{acknowledgments}

\end{document}